# Quasi-Periodic Nanoripples in Graphene Grown by Chemical Vapor Deposition and Its Impact on Charge Transport

*Guang-Xin Ni[1,2,3,*], Yi Zheng[1,2,*], Sukang Bae[4,*], Hye Ri Kim[4], Alexandre Pachoud[1,7], Young Soo Kim[4,9], Chang-Ling Tan[1], Danho Im[5], Jong-Hyun Ahn[4,6], Byung Hee Hong[4,8†], and Barbaros Özyilmaz[1,2,3,7‡]*

[1]Department of Physics, 2 Science Drive 3, National University of Singapore 117542

[2]NanoCore, 4 Engineering Drive 3, National University of Singapore, Singapore 117576

[3]Graphene Research Center, National University of Singapore, Singapore 117542

[4]SKKU Advanced Institute of Nanotechnology (SAINT) and Center for Human Interface Nano Technology (HINT), Sungkyunkwan University, Suwon 440-746, Korea

[5]Department of Chemistry, Sungkyunkwan University, Suwon 440-746, Korea

[6]School of Advanced Materials Science and Engineering, Sungkyunkwan University, Suwon 440-746, Korea

[7]NUS Graduate School for Integrative Sciences and Engineering (NGS), National University of Singapore, Singapore 117456

[8]Department of Chemistry, Seoul National University, Seoul, 151-747, Korea

[9]Department of Physics, Seoul National University, Seoul, 151-747, Korea

*These authors contributed equally to this work

†Corresponding authors: barbaros@nus.edu.sg, byunghee@snu.ac.kr

ABSTRACT The technical breakthrough in synthesizing graphene by chemical vapor deposition methods (CVD) has opened up enormous opportunities for large-scale device applications. In order to improve the electrical properties of CVD graphene grown on copper (Cu-CVD graphene), recent efforts have focussed on increasing the grain size of such polycrystalline graphene films to 100 micrometers and larger. While an increase in grain size and hence, a decrease of grain boundary density is expected to greatly enhance the device performance, here we show that the charge mobility and sheet resistance of Cu-CVD graphene is already limited within a single grain. We find that the current high-temperature growth and wet transfer methods of CVD graphene result in quasi-periodic nanoripple arrays (NRAs). Electron-flexural phonon scattering in such partially suspended graphene devices introduces anisotropic charge transport and sets limits to both the highest possible charge mobility and lowest possible sheet resistance values. Our findings provide guidance for further improving the CVD graphene growth and transfer process.

Graphene[1] is a promising material for many novel device applications such as ultrafast nanoelectronics, optoelectronics and flexible transparent electronics.[2-5] Cu-based CVD methods have now made wafer-scale graphene synthesis and transfer feasible both for single layer graphene[6,7](SLG) and bilayer graphene (BLG).[8] This not only brings the commercial applications of graphene within reach, but also provides great advantages in introducing new substrates to enhance and engineer its electronic properties by tuning the substrate-induced screening[9-12] and substrate-induced strain.[13,14] Unlike CVD graphene growth on Ni,[15,16] Cu-CVD graphene growth has a rather weak interaction with the underlying Cu substrate, allowing CVD graphene to grow continuously crossing atomically flat terraces, step edges, and vertices without introducing significant defects.[17] Thus, by controlling pre-growth annealing[7] and fine tuning growth parameters,[18,19] it is now possible to synthesize CVD graphene with sub-millimetre grain size. However, pre-growth annealing and CVD growth typically require high temperatures very close to the melting point of Cu at 1083 ºC. This leads to Cu surface reconstruction and local surface melting[17,20] during graphene growth, making high density Cu single-crystal terraces and step edges ubiquitous surface features. Taking into account the negative thermal expansion coefficient of graphene, this leads to new surface corrugations in CVD graphene during the cool down process.[21] Previously grain boundaries have been identified as one of the limiting factors to degrade graphene quality.[22] While the heptagon and pentagon network[22,23] at grain boundaries does disrupt the $sp^2$ delocalization of $\pi$ electrons in graphene, it remains to be seen whether this is indeed the charge scattering source most relevant for device applications. In this paper we show that Cu single-crystal step edges lead to the formation of quasi-periodic nanoripple arrays (NRAs) after transfer on $Si/SiO_2$ substrates. Such surface corrugations suspend up to 20 % of graphene and give rise to flexural phonon scattering. In particular at room temperature and density levels of the order of $10^{12}$ /cm$^2$ this leads to a strong anisotropy in the room-temperature (RT) conductivity depending on the relative orientation between NRA's and current flow direction. More importantly, flexural phonon scattering within the nanoripples sets a lower bound on the sheet resistance and upper bound on the charge carrier mobility even in the absence of grain boundaries.

RESULTS AND DISCUSSION

We first compare the RT resistivity vs gate voltage ($\rho$ vs $V_{BG}$) characteristics in four graphene field effect transistors (GFETs) of very different dimensions, ranging from the μm scale to the mm scale. The resistivities of these devices, fabricated from the same batch of CVD graphene, are presented in Fig. 1f. Surprisingly, except introducing stronger charge inhomogeneity, increasing the device channel area by 6 orders of magnitude does not significantly alter the charge carrier mobility; RT mobilities vary generally speaking independent of samples size between $\mu \sim$ 4000-6000 cm$^2$/Vs. This excludes grain boundaries (10−20 per mm, see Fig. 1a) as the main limiting factor for $\mu$ in our CVD graphene, and strongly suggests that the main scatterers are identical to the ones in exfoliated graphene: adatoms and/or charged impurities. Similar conclusions have recently been reached also by Ref [12].

However, high resolution contact mode atomic force microscopy (AFM) of CVD graphene on Si/SiO$_2$ with ultrasharp tips reveals a new type of surface corrugations (Fig. 2), whose influence on charge transport is not known. Distinct from the well known low density strain-induced wrinkles (~ 1 per 5 μm), we observe nanoripples of ~3±1 nm in height of much higher density (~ 10 per 5 μm), which are typically arranged in a quasi-periodic fashion (Fig. 2b). Each nanoripple location contains multiple peaks of 10-20 nm width (see Fig. 2e and SI), thus making it possible that overall a section of up to ~ 100 nm, *i.e.* up to 20 % of the graphene sheet becomes effectively suspended.

Systematic AFM studies on centimetre size samples further confirm that quasi-periodic nanoripples arrays (NRAs) are a general feature of the CVD graphene-on-SiO$_2$ surface morphology (See SI). To find out the origin of these NRAs, we carefully examined the single crystal surface of Cu substrates. As shown in Fig. 2a, thermally annealed Cu has a characteristically high density of single-crystal terraces and step edges. The terrace structure is typically a few hundred nanometres in width, separated by step edges of ~ 100 nm in width. These parameters agree very well with the dimensions of NRAs. The patterning of Au alignment mark arrays on Cu right after graphene growth allows the direct correlation of the local Cu step edge pattern, density and orientation with the surface morphology of graphene after it is transferred to a substrate. This comparison of graphene transferred from different Cu grains clearly demonstrates that NRAs originate from Cu step edges and thus, rules out any other factors during the transfer and fabrication process (See SI). Previous studies[6,7] proposed that the wrinkle formation, *i.e.* out-of-plane mechanical deformations in graphene sheets on Cu, is sufficient to release strain arising from the difference in thermal expansion coefficients between graphene and Cu. The observation of high density NRAs suggests, that strain is in fact mainly released by high-density Cu step edges. Below we will show that at RT these NRAs ultimately also set a lower bound on the sheet resistance $R_\square$ and an upper bound for the charge mobility µ even if all other extrinsic scatterings are eliminated.

We now focus on T-dependent electrical transport measurements. In micron size devices the QHE is well developed for both SLG and BLG, as shown in Fig. 3a&b, respectively. From an application point of view, the zero field measurements of σ vs n are more relevant. They show a pronounced sublinear behaviour, not only in CVD SLG but also in CVD BLG devices. The sublinearity is strongest at RT and diminishes gradually with decreasing temperature, as shown in Fig. 3c&d, respectively. This is best studied by plotting the T-dependent part of the resistivity instead and represents the key finding of our experiments. Our data reveal a superlinear T-dependent resistivity for T> 50 K. Remarkably, such a metallic behaviour is observed in both SLG and BLG (Fig 3e). Previous studies on supported exfoliated samples only reported such a T-dependent resistivity in SLG, while BLG samples did not show any T dependence away from the charge neutrality point (CNP).[24] Such behaviour for BLG is expected only for suspended samples, where the T-dependent contribution to ρ (n, T) scales as $T^2/n$ and is generally associated with electron-flexural phonon (FP) scattering.[25] Indeed, the high density NRAs effectively decouple up to 20 % of CVD graphene sheets from the substrate, activating low energy FP excitations in both SLG and BLG even when the samples are overall supported.

For CVD BLG, this clearly demonstrates that at RT NRAs will limit both $R_\square$ and $\mu$ due to FP scattering. However, the CVD SLG case is more ambiguous. Its resistivity has additional T dependent contributions due to scattering from remote interfacial phonons (RIP) of the SiO$_2$ substrate[26,27] and possibly high energy FPs arising from quenched 10 nm-wide nanoripples.[28,29] On SiO$_2$ substrates both the FP and RIP scattering mechanisms lead to a very similar T and n dependent behaviour over 50-350 K and ~10$^{12}$/cm$^2$ ranges (Fig. 3e&f).[27]

To explicitly measure the influence of NRAs on CVD SLG's resistivity, we have fabricated GFETs where the orientation of the electrodes is such that the current is either perpendicular ($\perp$) or parallel (//) to the NRAs (Fig. 4a&b). In total 4 sets of devices have been characterized. Here we discuss representative data for 2 sets (Namely S1, S2). We analyzed the corresponding transport data by assuming a resistivity $\rho$ of the form:

$$\rho(n, T) = \rho_0(n) + \alpha T + \rho_S(n,T) \qquad (1)$$

where $\rho_0(n)$ is the T-independent residual resistivity, $\alpha T$ is the acoustic phonon (AP) induced resistivity ($\alpha = 0.1 \Omega/K$),[27] and $\rho_S(n,T)$ the superlinear part of the resistivity. In Fig. 4c&e, we directly compare $\rho_S(n,T)$ for the $\perp$ and // devices by computing the increase in resistivity between 100K and T at fixed density n, namely $\Delta\rho_\perp(n, T) = \rho_\perp(n, T) - \rho_\perp(n, 100K) - \alpha(T-100K) = \rho_{S\perp}(n,T) - \rho_{S\perp}(n,100K)$ and $\Delta\rho_{//}(n, T) = \rho_{//}(n, T) - \rho_{//}(n, 100K) - \alpha(T-100K) = \rho_{S//}(n,T) - \rho_{S//}(n,100K)$. Strikingly in both samples, $\Delta\rho_\perp$ remains always significantly greater than $\Delta\rho_{//}$. In other words, the RT CVD graphene resistivity is anistropic. This is in sharp contrast with the isotropic resistivity of exfoliated samples and clearly shows that the phonon scattering rate is higher in the devices with the $\perp$ configuration. Since FPs are the only phonons which are activated upon suspension, this demonstrates that NRAs contribute also in CVD SLG importantly to the T-dependence of $\rho$.

Assuming a simple resistor-in-series and resistor-in-parallel model (See SI), we estimate the impact of NRAs on key figures of merit such as $\mu$ and $R_\square$ (Fig. 4d). Note that in our model $\rho_S(n,T)$ arises from both electron-FP scattering (in the nanoripples) and electron RIP-scattering events (between the nanoripples) independent of the NRAs orientation. With this we write $\rho_{S\perp} = f \cdot \rho_{FP} + (1-f) \cdot \rho_{RIP}$ and $\rho_{S//} = (f/\rho_{FP} + (1-f)/\rho_{RIP})^{-1}$, where f is the ratio of the typical ripple width w and the mean inter-ripple spacing a. Besides, we assume $\rho_{FP}$ is of the form $\gamma T^2/n$[29-31] and $\rho_{RIP}$ can be written as $(A/n) \cdot (g_1 /(\exp(E_1/(k_BT))-1) + g_2/(\exp(E_2/(k_BT))-1))$, where $g_1$ = 3.2 meV and $g_2$ = 8.7 meV are the respective coupling strengths of the $SiO_2$ RIP modes of energies $E_1$ = 63 meV and $E_2$ = 149 meV.[26,27] We can now estimate the two free parameters A and $\gamma$ setting the magnitude of $\rho_{FP}$ and $\rho_{RIP}$ by fitting the curves of Fig. 4c and 4e (See SI). This leads to A ~ $3\times10^{17}$ k$\Omega$/(eVcm$^2$), in reasonable agreement with Refs [26,27], and $\gamma$ ~ $6\times10^{-6}$ Vs/(mK)$^2$. Interestingly, these extracted values of $\gamma$ match well the experimental $\gamma$ values recently obtained for fully suspended graphene samples.[30] With this, it is now possible to predict FP-induced limits on CVD SLG's $\mu$ and $R_\square$. Fig. 4d shows the calculated RT mobility as a function of the Helium-T mobility $\mu_0$ for CVD graphene with f = 20 % both in $\perp$ and // orientations. As $\mu_0$ is unaffected by phonons, this is a convenient variable to gauge the influence of FPs.[32] Including AP scattering NRAs limit the RT mobility to ~40,000 cm$^2$/Vs in $\perp$ orientation and ~80,000 cm$^2$/Vs in // orientation, independent of the choice of substrate. In contrast, RT mobilities greater than 100,000 cm$^2$/Vs have already been achieved for exfoliated graphene encapsulated in h-BN.[33]

CONCLUSIONS

For large scale transparent electrode and display applications at RT, $R_\square$ is a more relevant number. Here, electron-FP scattering introduced by NRAs increases $R_\square$ unacceptably, given the industry requirement of $R_\square$ << 100 $\Omega$. At a technologically relevant carrier density of $1\times10^{12}$ /cm$^2$, the FP-induced increase in sheet resistance $\Delta R_\square$ alone is approximately 80 Ω independent of the AP and RIP scattering induced contributions to the sheet resistance $R_\square$ (Fig. 4d). Various approaches may be employed to overcome this issue. One may either try to reduce the effect of NRAs by inducing high charge carrier densities ($\Delta R_\square \sim f\gamma T^2/ne \sim 2$ Ω at $n = 5\times10^{13}$ /cm$^2$) and/or straining engineering,[25] or eliminating the NRAs altogether by transferring graphene under strain or using wet transfer processes which do not require any polymer coating. The rippling in CVD graphene can most likely never be fully avoided, but engineering Cu substrates properly may significantly reduce their presence (See SI).

In summary, we show that the current growth and transfer methods of CVD graphene lead to quasi-periodic nanoripple arrays in graphene. Such high density NRAs partially suspend graphene giving rise to flexural phonon scattering. This not only causes anisotropy in charge transport, but also sets limits on both the sheet resistance and the charge mobility even in the absence of grain boundaries. At room temperature NRAs are likely to play a limiting role also for the mobility of ultra-clean samples, in particular when the graphene sheets are transferred onto ultraflat BN substrates.[11,12] On the other hand the controlled rippling of graphene (SI) may be useful for graphene-based sensor applications as the ripples are more prone to adsorptions than flat graphene.[34] Controlled rippling may also be instrumental for spin-based device applications requiring surface modifications.[35]

METHODS

**CVD Graphene Synthesis.** The synthesis and transfer of large-scale CVD graphene are the same as in Ref. [7]. Electron backscattering diffraction reveals that the annealed Cu(001) substrates have single-crystal patches of Cu(111) and Cu(101), indicating local surface melting and recrystallization during growth (See Supplementary Information (SI)). The grain size of our CVD graphene is ~ 50−100 μm, as determined by scanning electron microscopy (SEM) of sub-monolayer graphene on Cu foil (Fig. 1a).

We can synthesize CVD graphene with a high BLG coverage up to 40% (Fig. 1b) or SLG dominant samples (> 95%).[18] Raman spectra (Fig. 1c) show insignificant defect peaks demonstrating the high quality of both SLG and A-B stacked BLG. Except for areas with optically visible wrinkles, Raman imaging with micrometer resolution also shows that on this scale strain is negligible (Fig. 1c and SI). Furthermore, Scanning Kelvin probe microscopy is used to confirm energy favourable A-B stacking structure[36] in CVD BLG (See SI).

**AFM Measurements.** Both, high resolution contact mode and tapping mode AFM technique have been utilized to characterize graphene morphology on top of copper and on top of the $Si/SiO_2$ substrate. For contact mode AFM ultra-sharp tips with radii as small as 10 nm were used limiting the error to ~ 10% error when measuring the 100 nm nanorippled area in Fig.2 and SFig.4. However, the contact mode AFM tips are more vulnerable to surface contaminations. Thus, for large-scale characterization tapping mode AFM was used.

**Raman Spectrascopy.** Raman spectroscopy/imaging were carried out with a WITEC CRM200 Raman system with 532 nm (2.33 eV) excitation and laser power at sample below 0.1 mW to avoid laser induced heating. A 100× objective lens with a NA=0.95 was used in the Raman experiments. To obtain the Raman images (see SI), a piezo stage was used to move the sample with step size of 200 nm and Raman spectrum was recorded at every point. The stage movement and data acquisition were controlled using ScanCtrl Spectroscopy Plus software from WITec GmbH, Germany. Data analysis was done using WITec Project software.

**Graphene FET Device Fabrication.** Graphene field effect transistor (GFET) Hall bars and four-terminal devices ranging in size from (1.2 × 0.8) $\mu m^2$ to (100 × 10) $\mu m^2$ were patterned by e-beam lithography (EBL) for metal contacts (5 nm Cr/30 nm Au) and $O_2$ plasma etching. Very large-scale GFETs of 1.2×1.2 $mm^2$ were etched into van der Pauw geometry by EBL followed by metal contacts evaporation using shadow masks. To precisely define fourcontacts either perpendicular or parallel with the NRAs, Au alignment mask arrays were pre-patterned using standard EBL processes followed by systermatic non-contact mode atomic force microscopy (AFM) scanning. The devices were finally thermally annealed at 400 K in high vacuum level ($10^{-5}$ mbar) for 2 hrs to clean the graphene working channel.

**Transport Measurements.** Electrical transport measurements were done in vacuum in a four contact configuration using a lock-in amplifier with an excitation current of 100 nA. T-dependent measurements were done from 350 K to 2 K in varable temperature insert (VTI) using standard four-contact lock-in techniques. In total 8 SLG devices and 3 BLG devices have been measured. Here we discuss 5 (2) representative SLG (BLG) devices in more detail.

*Acknowledgment:* We thank Chun-Xiao Cong and Ting Yu for Raman characterization of Cu-CVD graphene, Xiang-Ming Zhao, Kai-Wen Zhang, Xiang-Fan Xu, Chee-Tat Toh for assisting in device fabrications, and Manu Jaiswal for useful discussions. This work is supported by the Singapore National Research Foundation grant NRF-RF2008-07, NUS/SMF grant, US Office of Naval Research (ONR and ONR Global) and A*STAR SERC TSRP-Integrated Nano-photo-Bio Interface (R-144-000-275-305), NUS NanoCore, and by Basic Science Research Program (2011K000615, 2011-0017587, 2011-0006268) and Global Research Lab. (GRL) Program (2011-0017587) through the National Research Foundation of Korea (NRF) funded by the Ministry of Education, Science, and Technology.

*Supporting Information Available:* The contents of Supporting Information include the following: (1) Ultra-large copper grain size by pre-growth annealing, (2) SKPM of CVD grown graphene on Cu, (3) Correlation of structure of Cu and CVD graphene after wet transfer, (4) CVD graphene transferred on Si/SiO2 substrate without any other supporting., (5) How to avoid nanoripple arrays?, (6) T-dependent sublinear conductivity in CVD BLG FETs, (7) Raman spectroscopy 2-D mapping of CVD SLG device, (8) Modelling of the anisotropic resistivity, (9) Method used to extract A and $\gamma$ for $\rho_{RIP}$ and $\rho_{FP}$, (10)

Anisotropic resistivity as a function of charge density for sample set S1. This material is available free of charge *via* the Internet at http://pubs.acs.org.

FIGURE CAPTIONS

**Fig. 1**. (a) SEM image of sub-monolayer graphene on Cu foil. (b) Optical image of high BL coverage CVD graphene on 300 nm $SiO_2$. The black, red and blue circles indicate the Raman measurement (0.4 μm in spot size) locations. Black arrows indicate wrinkles formed during growth. (c) Raman spectroscopy of SLG, BLG and multilayer graphene. Raman on optical visible wrinkle shows significant broadening in G and 2D peaks, indicating non-negligible strain. (d) Optical image of an mm size GFET with van der Pauw geometry and (e) of μm size. Scale bar: 2 μm. (f) Electrical measurements of $mm^2$ and $μm^2$ devices at RT. Each curve is shifted by + 20 V. Inset show the corresponding $\sigma-\sigma_{min}$ vs n.

**Fig. 2**. (a) AFM image of Cu surface, showing single crystal terraces and step edges. Color scale: 0 ~ 60 nm. Scale bar: 1 μm. (b) and of CVD graphene on $SiO_2$, showing high density nanoripples induced by Cu step edges. Color scale: 0 ~ 15 nm. Scalebar: 1 μm. (c) Upper part: AFM cross-section of Cu terraces. The typical width of step edges is ~ 100 nm. Lower part: AFM line scans of graphene after transfer to $Si/SiO_2$ reveal nanoripple arrays which are closely correlate with the Cu terraces. (d) Illustration of nanoripple formation, structure and periodicity. (e) High resolution scan with ultrasharp tips show that nanoripple consists of multiple peaks of 10-20 nm width.

**Fig. 3**. (a) and (b) QHE of CVD SLG and BLG graphene on $Si/SiO_2$ substrate, respectively. SLG shows the anomalous quantization plateaux of $\pm 4e^2/h(N+1/2)$, while BLG has the typical $\pm 4Ne^2/h$ quantization signatures. (c) T-dependent sub-linear behaviour of a SLG. Inset: Insulating behaviour with $n < 5 \times 10^{11}$ /$cm^2$. (d) T-dependent sub-linear behaviour of a BLG. Inset: AFM image of a flower shaped CVD BLG, quasi-periodic NRAs are clearly seen in both SLG and BLG, the scale bar is 1 μm. (e) $\rho(T)$ for SLG at $n = 0.5 \times 10^{12}$ /$cm^2$ and $n = 2.8 \times 10^{12}$ /$cm^2$ and for BLG at $n = 0.5 \times 10^{12}$ /$cm^2$ and $n = 3.8 \times 10^{12}$ /$cm^2$. Dashed lines correspond to a two-parameter fit to the data using $\rho = \rho_0 + 0.1T + \gamma T^2/ne$, and serve as guide to the eyes. (f) RT superlinear contribution $\rho_S$ to the total resistivity $\rho$ as a function of charge density n for CVD SLG FET (black squares) and CVD BLG FET (white squares). The solid red and blue curves correspond to fits of the form $\rho_S = a/n$, where $\rho_S$ arises from both FPs and RIPs. We used a = $1.57 \times 10^{18}$ $\Omega/m^2$ and $2.87 \times 10^{18}$ $\Omega/m^2$ for SLG and BLG, respectively.

**Fig. 4**. (a-b) T-dependent sub-linear conductivity for both $\perp$ and // NRAs configurations. Inset: AFM image of graphene channel with clear $\perp$ and // NRAs orientations. The scale bar is 1 μm. (c) Anisotropic resistivity results obtained with sample set S1 at $n = 2 \times 10^{12}$ /$cm^2$ (for data obtained at higher carrier

density, see SI); The blue and red data points correspond to the $\perp$ and $//$ NRAs configurations, respectively. They are fitted with $\rho_{S\perp}(n,T) - \rho_{S\perp}(n,100K)$ (solid blue) and $\rho_{S//}(n,T) - \rho_{S//}(n,100K)$ (solid red). The black points correspond to $\Delta\rho_{inter}$, and are extracted from $\Delta\rho_\perp$ and $\Delta\rho_{//}$ following the method described in SI. $\Delta\rho_{inter}$ is fitted with $\rho_{RIP}(A, n, T) - \rho_{RIP}(A, n, 100K)$ with $A \sim 3\times10^{17}$ k$\Omega$/eVcm$^2$ (black solid curve). The inset shows the optical image of the device set. The scale bar is 10 μm. (d) Estimate of the NRA impact on CVD SLG R□ and μ at $n = 2 \times 10^{12}$ /cm$^2$. The solid red and black curves represent the RT mobility μ against liquid Helium T mobility $\mu_0$ for CVD SLG in both $\perp$ (solid red) and $//$ (solid black) NRAs configurations, assuming electron-RIP scattering is suppressed. The dashed dotted curves represent the FP-induced increase in R□ in $\perp$ (red) and $//$ (black) orientations. (e) $\Delta\rho(T)$ obtained from a second set of devices S2 for different charge densities ranging from $2.0 \times 10^{12}$ /cm$^2$ to $3.4 \times 10^{12}$ /cm$^2$. The data points and fitting curves have the same definitions as described in (c).

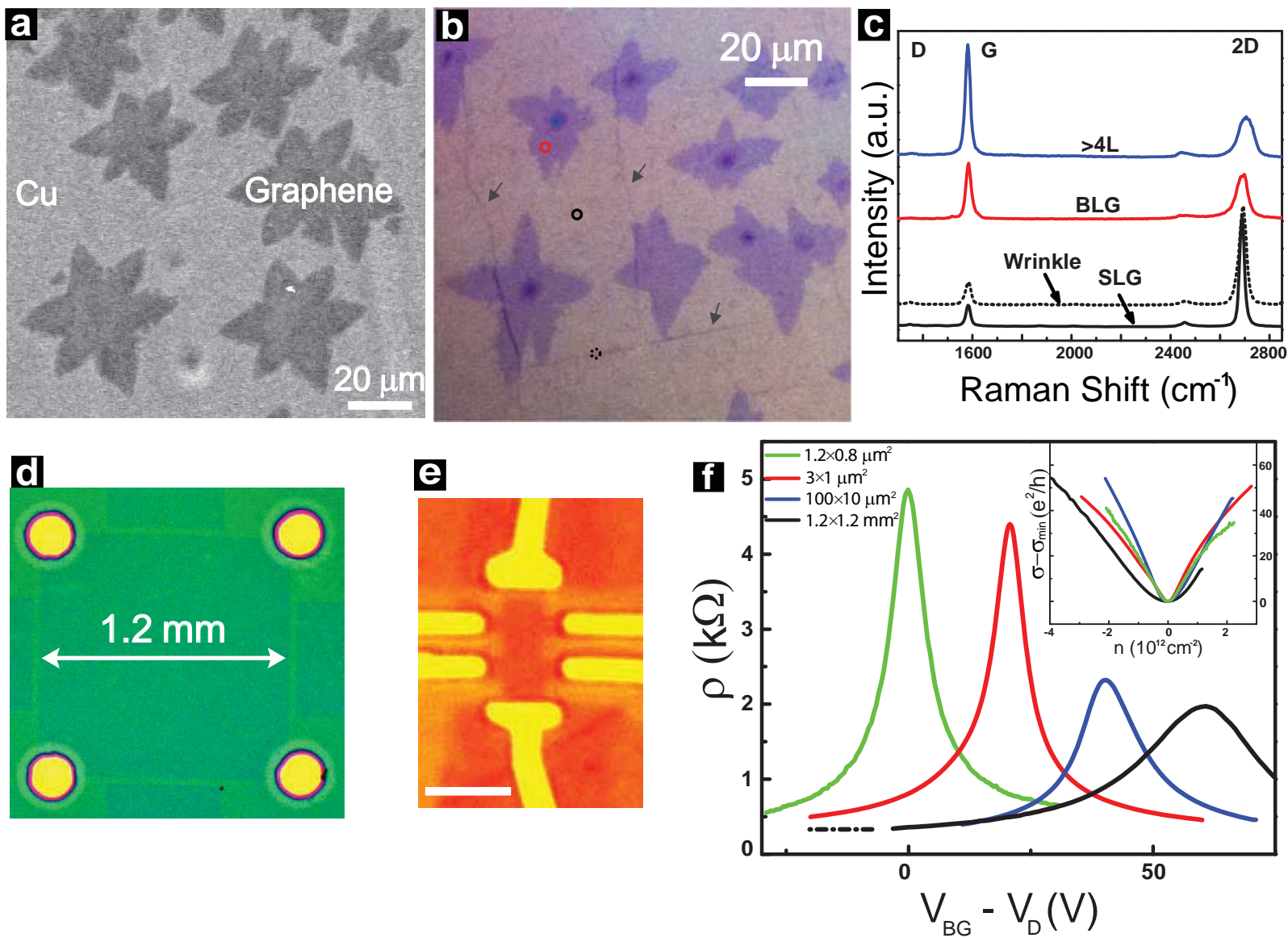

Fig. 1 Guang-Xin Ni et al.

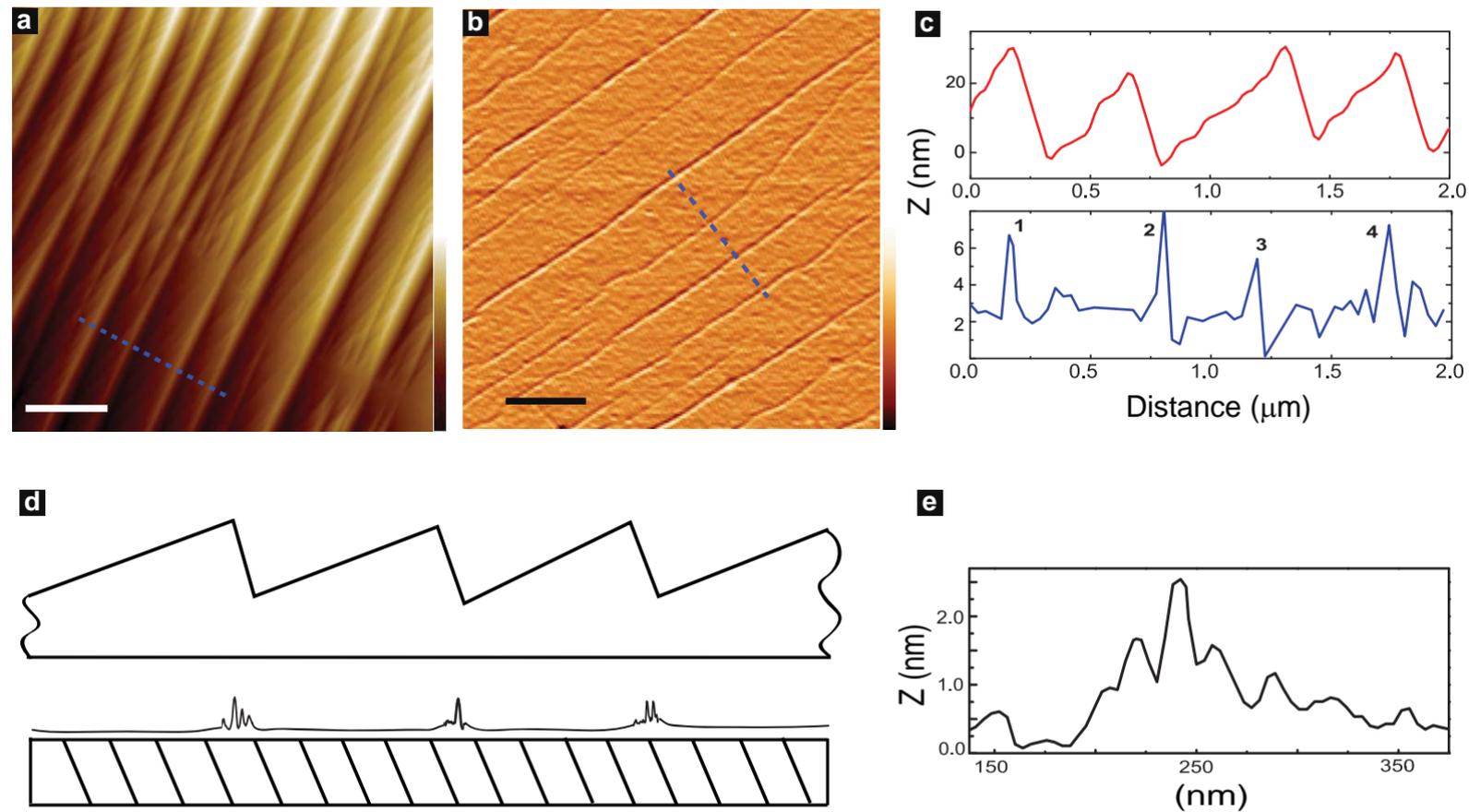

Fig. 2 Guang-Xin Ni et al.

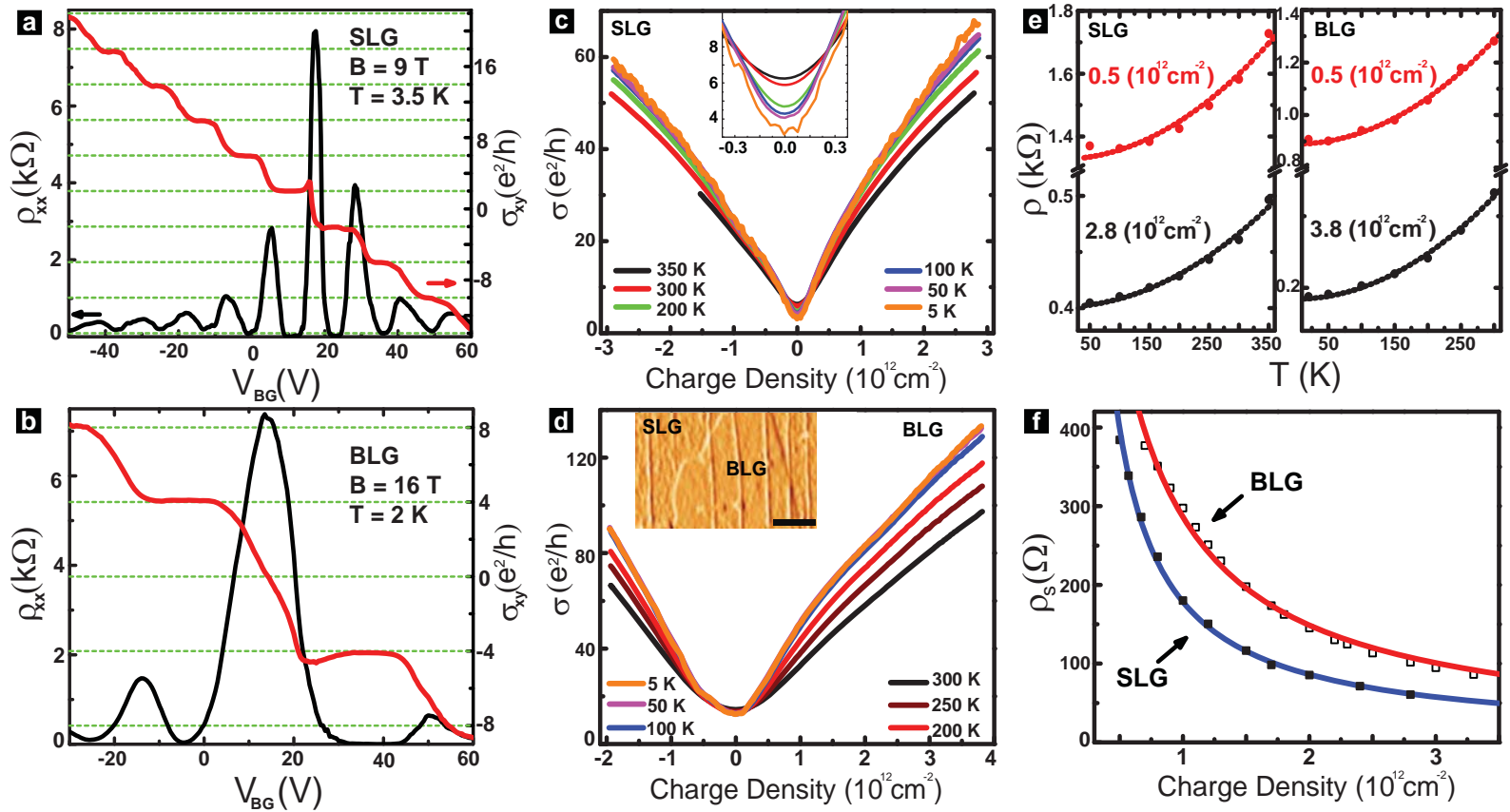

Fig. 3 Guang-Xin Ni et al.

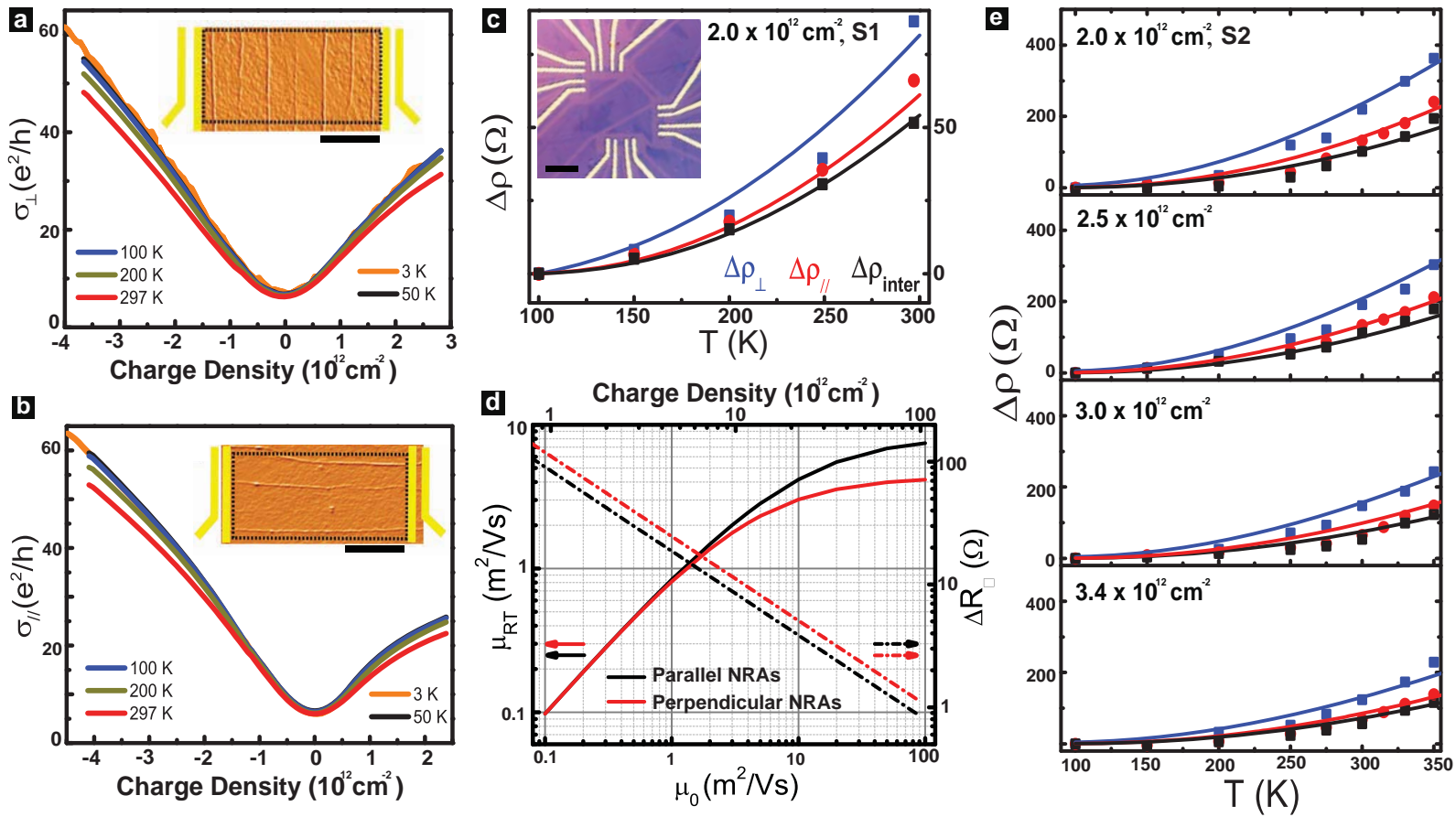

Fig. 4 Guang-Xin Ni et al.